\documentclass{article}
\usepackage{graphicx}
\usepackage{amsmath}
\usepackage{dcolumn}
\usepackage{bm}
\DeclareMathAlphabet{\mathpzc}{OT1}{pzc}{m}{it}

\newcommand{\pde}[2]{\frac{\partial {#1}}{\partial {#2}}}

\newcommand{\matd}[1]{\frac{\text{D}{#1}}{\text{D}t}}

\begin{document}
\title{Compressible flow in a Noble-Abel Stiffened-Gas fluid}
\author{
M. I. Radulescu\\
Department of Mechanical Engineering\\
University of Ottawa, Ottawa (ON) K1N 6N5 Canada\\
}

\date{\today}

\maketitle
\begin{abstract}
While compressible flow theory has relied on the perfect gas model as its workhorse for the past century, compressible dynamics in dense gases, solids and liquids have relied on many complex equations of state, yielding limited insight on the hydrodynamic aspect of the problems solved.  Recently, Le Métayer and Saurel studied a simple yet promising equation of state owing to its ability to model both the thermal and compressibility aspects of the medium.  It is a hybrid of the Noble-Abel equation of state and the stiffened gas model, labeled the Noble-Able Stiffened Gas (NASG) equation of state.  In the present work, we derive the closed form analytical framework for modelling compressible flow in a medium approximated by the NASG equations of state.  We derive the expressions for the isentrope, sound speed, the isentropic exponent, Riemann variables in the characteristic description, and jump conditions for shocks, deflagrations and detonations. We also illustrate the usefulness by addressing the Riemann problem.  The closed form solutions generalize in a transparent way the well-established models for a perfect gas, highlighting the role of the medium's compressibility.  
\end{abstract}

\section{Introduction}
Recenty, Le Métayer and Saurel (henceforth LS) have analyzed in detail an equation of state consisting of a blend of the Noble-Abel equation of state and the stiffened gas equation of state, labeled the Noble-Abel Stiffened Gas (NASG) \cite{lemetayer2016}.  This equation of state has proven quite successful in numerical work treating compressible flows of multi-phase and multi-component flows, both inert and reactive\cite{furfaro2019towards, boivin2019}. It has also been shown to empirically capture very well the compressibility of metals,  liquids and dense gases \cite{richards1923compressibility, bridgman1924}.

The model generalizes the model usually referred to in the modern literature as the \textit{modified Tait}, \textit{Tammann} \cite{ivings1998} (although Tammann postuated the present form of the NASG model \cite{tammann1912}) or \textit{stiffened gas} \cite{menikoff2007}, which has an ill-defined thermodynamic temperature when $\gamma$ in \eqref{eq:eNASG} is artificially increased to account for the correct compressibility of the medium- see Radulescu \cite{radulescu2019} for discussion.  It includes the co-volume factor present in the Noble-Abel equation of state, which was present in the early work of Tammann.  This permits to recover the correct compressibility while also capturing the thermal behaviour. 

 In the present study, we would like to show that the NASG equation of state also permits to analytically tackle problems of compressible inert or reactive flows quite simply.  We extend the results of LS and derive the relevant quantities to treat compressible flows,  i.e., the isentrope, sound speed, Riemann variables in the characteristic form of the Euler equations and their weak solutions for shocks, deflagrations and detonation waves.  We generalize the well accepted results of a perfect gas; the NASG model offers the same simplicity and power to tackle problems in dense gases, solids and liquids.

\section{The isentrope, sound speed and the isentropic exponent}
The NASG equation of state for a single component relates the internal energy $e$ of the medium to the medium's pressure $p$ and specific volume $v$:
\begin{equation} \label{eq:eNASG}
e(p,v)=\frac{p+\gamma p_\infty}{\gamma-1}\left(v-b \right)+q
\end{equation}
where $p_\infty$, $b$ and $q$ are fitting parameters and $\gamma$ is the ratio of specific heats.  The physical meaning of $b$ is the usual co-volume, or minimum effective volume that the fluid can occupy given its finite physical size.  The parameter $p_\infty$ accounts for the attraction forces between the molecules, such that gas like behaviour is only expected for pressures significantly larger than $p_\infty$.  

LS and Radulescu \cite{radulescu2019} derived the corresponding temperature equation of state starting from \eqref{eq:eNASG}: 
\begin{equation}\label{eq:TNASG}
T=\frac{(p+ p_\infty)(v-b)}{C_v(\gamma-1)}
\end{equation}
The right hand side is the functional form for an isotherm postulated by Tammann \cite{tammann1912} and subsequently used to model dense gases, liquids and solids \cite{richards1923compressibility, bridgman1924}.      

To obtain the insentrope, sound speed and the isentropic exponent, we begin by the first Gibbs, or $TdS$ equation, given by
\begin{equation}\label{eq:1-74}
de=Tds-pdv
\end{equation}
which we re-write in terms of density $\rho=1/v$ as
\begin{equation}\label{eq:1-81}
de=Tds+\frac{p}{\rho^2}d\rho
\end{equation}
or as a perfect differential
\begin{equation}\label{eq:1-82}
de(p, \rho)=\left(\pde{e}{p}\right)_\rho dp+\left(\pde{e}{\rho}\right)_p d\rho
\end{equation}
Equating both expressions to eliminate $de$ we obtain
\begin{equation}\label{eq:1-83}
dp=\frac{\frac{p}{\rho^2}-\left(\pde{e}{\rho}\right)_p}{\left(\pde{e}{p}\right)_\rho}d\rho+\frac{T}{\left(\pde{e}{p}\right)_\rho}ds
\end{equation}
Now write $p(\rho, s)$ as a perfect differential of the form 
\begin{equation}\label{eq:1-84}
dp=\left(\pde{p}{\rho}\right)_sd\rho+\left(\pde{p}{s}\right)_\rho ds
\end{equation}
and comparing \eqref{eq:1-83} with \eqref{eq:1-84}, we get immediately the expression for the sound speed
\begin{equation}
c^2\equiv\left(\pde{p}{\rho}\right)_s=\frac{\frac{p}{\rho^2}-\left(\pde{e}{\rho}\right)_p}{\left(\pde{e}{p}\right)_\rho}\label{eq:1-85}
\end{equation}
Evaluating the partial derivatives $\left(\pde{e}{\rho}\right)_p$ and $\left(\pde{e}{p}\right)_\rho$ from \eqref{eq:eNASG}, we obtain immediately the expression for the sound speed in a NASG fluid:
\begin{equation}\label{eq:sound}
c^2=\gamma \frac{p+p_\infty}{\rho(1-\rho b)}
\end{equation}
and similarly using \eqref{eq:TNASG}, we can re-write \eqref{eq:1-83} as  
\begin{equation}\label{eq:ds}
dp=c^2 d\rho + \left(p+p_\infty \right) \frac{ds}{C_v} 
\end{equation}
On the isentrope, $ds=0$, and \eqref{eq:ds} integrates to:
\begin{equation}\label{eq:isentrope}
(p+p_\infty)(v-b)^\gamma=const. 
\end{equation}
We thus recover the result for a perfect gas with $p+p_\infty$ replacing $p$ and $v-b$ replacing $v$.   Note that for $b$=0, the isentrope is what is referred to in the literature as the Tait equation of state, popularized by Kirkwood and Bethe \cite{kirkwood1942}. 

The isentropic exponent becomes
\begin{equation} \label{eq:isentropicexponent}
\gamma_s \equiv \left(\pde{\ln p}{\ln \rho}\right)_s=\frac{\rho}{p}c^2=  \gamma \frac{1+\frac{p_\infty}{p}} {1- \frac{b}{v} }
\end{equation}
It highlights how the compressibility of a medium departs from that of a perfect gas when the pressure is comparable or less than the  stiffening pressure and/or when the volume approaches the minimum co-volume.  
 
\section{Characteristic description of quasi-1D flows}
To appreciate the usefulness of the NASG equation of state to describe the general motion of a compressible fluid, we first derive the general formulation for an arbitrary equation of state.  The mass and momentum conservation for a quasi-1D flow in a stream tube of varying cross-section, in the presence of a body force $f$ are
\begin{align}
\matd{\rho}+\rho\pde{u}{x}&=-\rho u \frac{1}{A}\pde{A}{x}\\
\rho\matd{u}&=-\pde{p}{x}+f
\end{align}
These are of course supplemented by a statement for conservation of energy along a particle path, for example \eqref{eq:1-83} written in general form: 
\begin{equation}
\matd{p}=c^2 \matd{\rho}+T \left(\pde{p}{e}\right)_\rho \matd{s}
\end{equation}  
Eliminating the density derivative and taking linear combinations of the remaining two equations, we can obtain the two characteristic equations:  
\begin{equation}
\frac{1}{\rho c}\matd{_\pm}p\pm \matd{_\pm}u\\
=- u c \pde{\ln A}{x}+\frac{T}{\rho c}  \left(\pde{p}{e}\right)_\rho\matd{s}\pm \frac{f}{\rho} \label{eq:5-42}
\end{equation}
where 
\begin{equation}
\matd{_\pm}=\pde{}{t}+(u\pm c)\pde{}{x}
\end{equation}
are the convective derivatives along the $C_\pm$ characteristic directions $dx/dt= u\pm c$. 

So far, relation \eqref{eq:5-42} applies to any medium. If the flow is isentropic, all variables appearing in the first term can be expressed in terms of a single thermodynamic variable.  After some manipulations using the isentrope and the form of the sound speed above, the first term can be re-written as:
\begin{equation}
\frac{1}{\rho c}dp= d \left(\frac{2}{\gamma-1}\sqrt{\gamma(p+p_\infty)(v-b)} \right)
\end{equation}
such that the characteristic equation for isentropic flow becomes 
\begin{align}
\matd{_\pm} J_\pm =- u c \pde{\ln A}{x}+\pm \frac{f}{\rho} \label{eq:characteristicequation}
\end{align}
where the Riemann variables $J_\pm$ for a NASG fluid are quite simply:
\begin{equation}
J_\pm= \left(\frac{2}{\gamma-1}\sqrt{\gamma(p+p_\infty)(v-b)} \right) \pm u \label{eq:J}
\end{equation}

Clearly, this reduces to the classic form $2c/(\gamma-1)\pm u$ in a perfect gas, for which $c^2=\gamma p v$.  For isentropic flows without lateral divergence and body forces, simple wave solutions can be obtained in the usual way by exploiting the constancy of one of the Riemann variables everywhere in the flow \cite{whitham}.  Simple wave extensions for weak shocks can be exploited in the usual manner \cite{whitham}. 

Non-isentropic flow, for example reactive flow, flows with strong shocks or with generic heat addition or losses, can be treated as well by involving thermodynamic derivatives of the thermodynamic part of the Riemann variable $\theta \equiv \frac{2}{\gamma-1}\sqrt{\gamma(p+p_\infty)(v-b)}$.
\begin{equation}
dp(\theta,S)=\left(\pde{p}{\theta}\right)_s d\theta+\left(\pde{p}{s}\right)_\theta ds
\end{equation}
This permits to re-write \eqref{eq:5-42} in terms of the Riemann variables and entropy changes along particle paths and $C_\pm$ characteristics. 

\section{Inert shock waves}

The jump conditions across shock waves also yield very simple expressions for a NASG fluid.  The weak form of the conservation laws of mass, momentum and energy across a shock wave moving at speed $D$ normal to its surface, into gas moving at speed $u_1$ normal to the shock take the usual form \cite{whitham}. 
\begin{align}
\frac{D-u_1}{v_1}=\frac{D-u_2}{v_2} \label{eq:massjump}\\
p_1+\frac{(D-u_1)^2}{v_1}=p_2+\frac{(D-u_2)^2}{v_2} \label{eq:momjump}\\
e_1+p_1v_1+\frac{1}{2}(D-u_1)^2=e_2+p_2v_2+\frac{1}{2}(D-u_2)^2 \label{eq:energyjump}
\end{align}
The post shock state is indicated with subscript 2.  It is simpler to work with the mass conservation \eqref{eq:massjump} and  linear combinations with the two other equations, leading to the so-called Hugoniot curve and Rayleigh line. The former is obtained by eliminating the speeds from the three equations, yielding 

\begin{align}
e_2-e_1=\frac{1}{2}(p_1+p_2)(v_1-v_2)\label{eq:hugoniot}
\end{align}
The Rayleigh line is the combination of the mass and momentum used to eliminate the speed $u_2$, yielding
\begin{align}
p_2-p_1=\frac{(D-u_1)^2}{v_1}\left(1-\frac{v_2}{v_1}\right) \label{eq:rayleigh}
\end{align}
The mass conservation \eqref{eq:massjump}, the Rayleigh line \eqref{eq:momjump} and the Hugoniot curve \eqref{eq:hugoniot} now provide the system of equations that determine the jump conditions.  These are general for any medium.  

Upon substitution of the equation of state expressions for a NASG fluid \eqref{eq:eNASG} for $e_1$ and $e_2$, the Hugoniot relation \eqref{eq:hugoniot} can be written after some manipulations:
\begin{equation}
\left(\frac{\overline{p}_2}{\overline{p}_1}+\frac{\gamma-1}{\gamma+1}\right) \left(\frac{\overline{v}_2}{\overline{v}_1}-\frac{\gamma-1}{\gamma+1}\right)=1-\left(\frac{\gamma-1}{\gamma+1}\right)^2
\end{equation}
where $\overline{p}=p+p_\infty$ and $\overline{v}=v-b$.  This is the same expression as for the Hugoniot curve for a perfect gas, $\overline{p}$ replacing $p$ and $\overline{v}$ replacing $v$.

Noting that the square of the shock Mach number can be written as
\begin{equation}
M_s^2 \equiv \frac{(D-u_1)^2}{c_1^2}=\frac{(D-u_1)^2 \overline{v}_1 }{\gamma \overline{p}_1 v_1^2}
\end{equation}
The Rayleigh line takes again the same form as that for a perfect gas, with $\overline{p}$ replacing $p$ and $\overline{v}$ replacing $v$:
\begin{equation}
\frac{\overline{p}_2}{\overline{p}_1}=1+\gamma M_s^2 \left(1- \frac{\overline{v}_2}{\overline{v}_1}\right)
\end{equation}

Solving for the pressure and specific volume jumps by satisfying the Rayleigh and Hugoniot conditions yield again the same expressions as for a perfect gas with $\overline{p}$ replacing $p$ and $\overline{v}$ replacing $v$:
\begin{align}
\frac{\overline{p}_2}{\overline{p}_1}=1+\frac{2 \gamma \left(M_s^2-1   \right)}{\gamma+1}\\
\frac{\overline{v}_2}{\overline{v}_1}=\frac{(\gamma-1)M_s^2+2}{(\gamma+1)M_s^2}
\end{align}
The change in particle speed across the shock wave follows from the specific volume jump and the conservation of mass \eqref{eq:massjump}, which, after some manipulations, yields
\begin{align}
\frac{u_2-u_1}{c_1}=\left(1-\frac{b}{v_1} \right)\frac{2(M_s^2-1)}{(\gamma+1)M_s}\label{eq:shock_speed}
\end{align}
This expression is also the same as for a perfect gas with the addition of the term $\left(1-\frac{b}{v_1} \right)$ accounting for the reduction in the medium's compressibility as $v_1\rightarrow b$, as already noted when discussing the isentropic exponent.

It is now clear that other jump conditions of interest, such as temperature, sound speed, flow Mach number, Riemann variables, entropy, etc... are easily obtained.

\section{The Riemann problem}
\begin{figure}
\begin{center}
\includegraphics[width=0.7 \columnwidth]{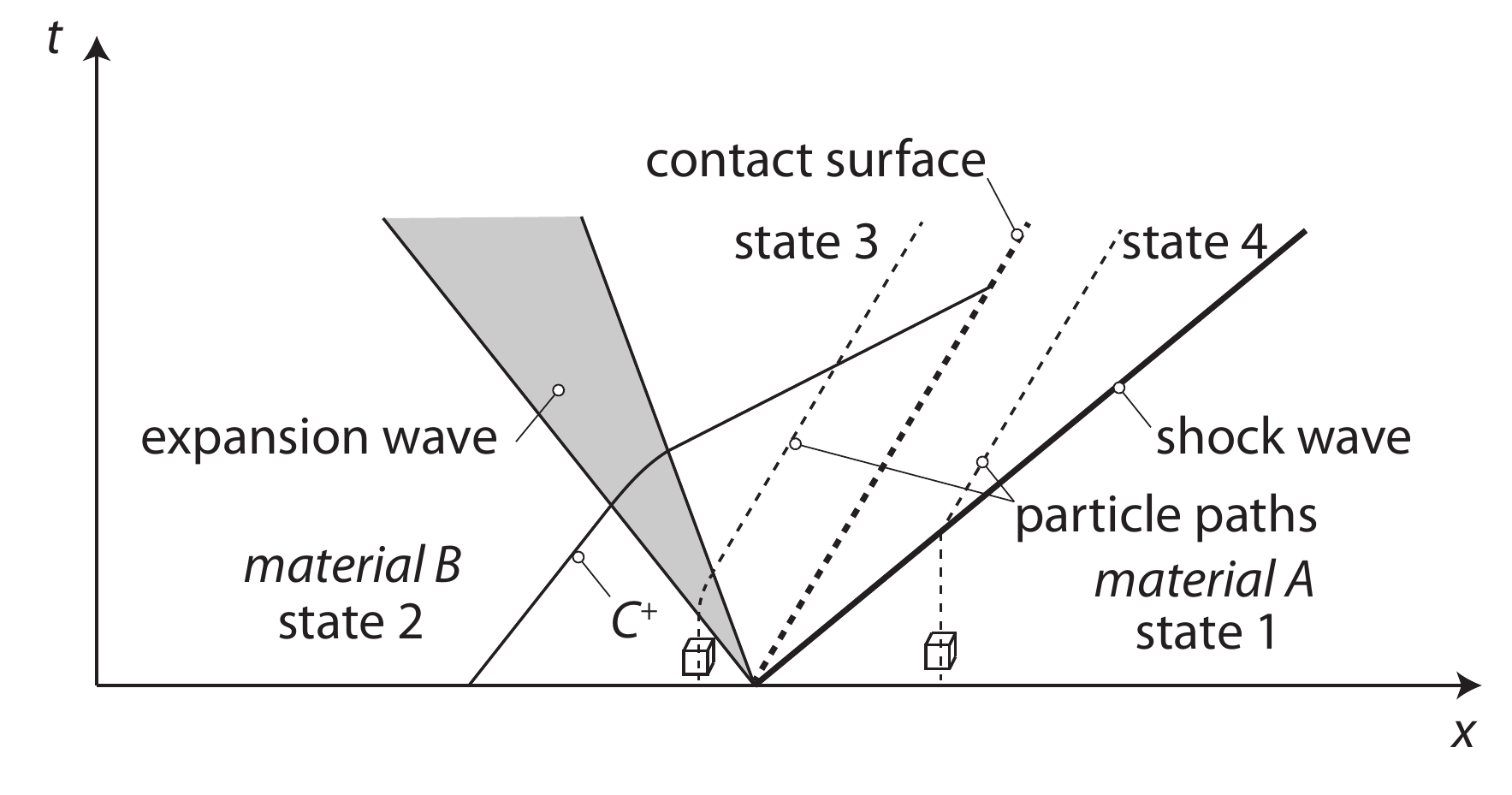}
\caption{The solution to the Riemann problem for $p_2 > p_1$ and $u_1 =u_2 =0$ for the mechanical equilibration at the interface between two different materials.}
\label{fig_Riemann}
\end{center}
\end{figure}

To illustrate the usefulness of the analytical treatment in a NASG medium, we can consider for example the solution to the initial value problem of two separate media initially at different but constant mechanical and thermodynamic states.  This is known as the Riemann problem.  Its importance is primarily in numerical work at grid cell interfaces (see, for example, Gottlieb and Groth's summary  \cite{gottlieb1988}).  Consider a medium \textit{A} at uniform state $(p_1, v_1, u_1)$, separated from a medium \textit{B} at state $(p_2, v_2, u_2)$.  The properties may be different in the two media, labeled respectively with subscripts \textit{A} and \textit{B}.  The general case is allowing $u_1 \neq u_2$ and $p_1\neq p_2$. Five different wave patterns are possible, involving either shock or expansion waves driven into each respective medium, one of which is shown schematically in the space-time diagram of Fig.\ \ref{fig_Riemann}.  While the general case is clearly relevant for numerical work in multi-material situations, and will be communicated in a sequel, we will focus here on the particular case $u_1 =u_2 =0$ of Fig.\ \ref{fig_Riemann} for illustrative purposes.  This is known as the \textit{shock tube problem}, since its solution provides an idealized solution of a high pressure gas initially at rest discharging into a low pressure one, driving a shock wave. 
When $p_2 > p_1$, a shock wave will be driven into medium \textit{A}, changing its state from \textit{1} to \textit{4} in Fig. \ref{fig_Riemann}.  The compression of medium \textit{A} is associated with the expansion of medium \textit{B}, whose state changes from \textit{2} to \textit{3}.  The solution to this problem is determining the strength and structure of the expansion wave and the strength of the shock wave that give rise to mechanical equilibrium at the interface (or contact surface) between the two media, i.e., $p_3=p_4$ and  $u_3 = u_4$.

A $C^+$ characteristic connects state \textit{2} to state \textit{3}.  Since the expansion of the medium along a particle path is also isentropic, state \textit{3} is linked to state \textit{2} by the invariance of the Riemann variable $J_+$ given by \eqref{eq:J} and the isentropic expansion condition \eqref{eq:isentrope}.  These two expressions can be combined to eliminate the specific volume at state 3, yielding:
\begin{equation}\label{eq:acrossexpansion}
 \frac{2}{\gamma_B-1}\sqrt{\gamma(p_2+p_{\infty,B})(v_2-b_B)} 
 \left( 
 1-\left(  \frac{p_3+p_{\infty,B}}{p_2+p_{\infty,B}}\right)^{\frac{\gamma_B-1}{2\gamma_B}}  
 \right) - u_3 =0
\end{equation}
The unknown pressure $p_3$ and particle speed $u_3$ satisfy the mechanical equilibrium condition $p_3=p_4$ and $u_3 = u_4$.  Making these substitutions in \eqref{eq:acrossexpansion} results in a condition linking $p_4$ and $u_4$.  Since these satisfy the shock jump equations derived in the previous section in terms of the shock Mach number $M_s$, namely
\begin{equation}
\frac{p_4+p_{\infty,A}}{p_1+p_{\infty,A}}=1+\frac{2 \gamma_A \left(M_s^2-1   \right)}{\gamma_A+1}
\end{equation}
and
\begin{equation}
\frac{u_4}{c_1}=\left(1-\frac{b_A}{v_1} \right)\frac{2(M_s^2-1)}{(\gamma_A+1)M_s}
\end{equation}
we have obtained a single algebraic condition for $M_s$.  Given the shock and expansion wave strengths, the structure of the expansion wave is also found with minor effort in closed form using the simple wave argument \cite{whitham}.  The shock tube problem treated for two different materials applies evidently to the case when either or both of the media are ideal gases, by setting the coefficients of $b$, $\gamma$ and $p_{\infty}$ accordingly.  

\section{Deflagrations and detonations}
\begin{figure}
\begin{center}
\includegraphics[width=0.7 \columnwidth]{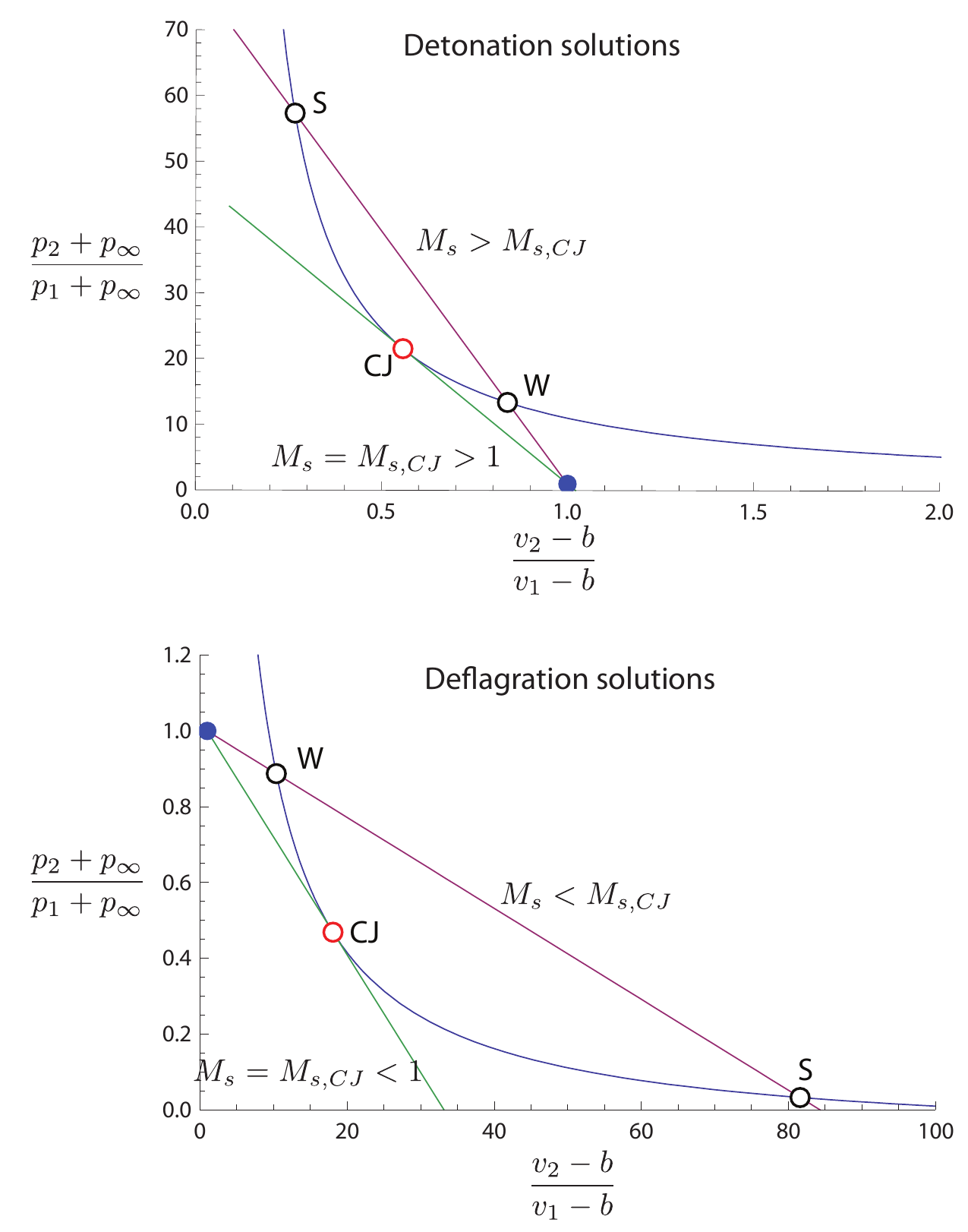}
\caption{Weak (W), strong (S) and Chapman-Jouguet (CJ) solutions for detonation and deflagrations at the intersection of the Rayleigh lines (purple) with the Hugoniot curve (blue) for $\frac{Q}{\overline{p}_1\overline{v}_1}=50$ and $\gamma=1.2$; the CJ Rayleigh line is green and the initial state is blue. }
\label{fig1}
\end{center}
\end{figure}

Gasdynamic reactive discontinuities in a NASG fluid can also be quite simply modelled by extensions to the well-established perfect gas model.  Here we seek the possible discontinuities with energy addition or withdraw by simply changing the constants $q_1$ and $q_2$ accounting for changes in reference internal energy of the medium ahead and behind the wave.  If we let the effective heat release across the wave be $Q=q_1-q_2$, proceeding as for the inert shocks treated above, the Hugoniot curve given by \eqref{eq:hugoniot} becomes:
\begin{equation}
\left(\frac{\overline{p}_2}{\overline{p}_1}+\frac{\gamma-1}{\gamma+1}\right) \left(\frac{\overline{v}_2}{\overline{v}_1}-\frac{\gamma-1}{\gamma+1}\right)
=1-\left(\frac{\gamma-1}{\gamma+1}\right)^2 +2\frac{\gamma-1}{\gamma+1}\frac{Q}{\overline{p}_1\overline{v}_1}
\end{equation} 
This is again the same Hugoniot expression as for a perfect gas, with  $\overline{p}$ replacing $p$ and $\overline{v}$ replacing $v$.  The Hugoniot curve is shown in Fig.\ \ref{fig1} in the compression region for detonations ($v_2<v_1$) and in the expansion region ($v_2 > v_1$) for deflagrations.

The Rayleigh line expression is the same as for an inert shock.  Its combination with the Hugoniot curve readily permits to obtain the jump equations for the pressure and specific volume:
\begin{align}
\frac{\overline{p}_2}{\overline{p}_1}=\frac{1+\gamma M_s^2}{\gamma+1} \mp \gamma M_s^2 \sqrt{\zeta}\\
\frac{\overline{v}_2}{\overline{v}_1}=\frac{1+\gamma M_s^2}{M_s^2 (\gamma+1)} \pm \sqrt{\zeta}
\end{align}
where 
\begin{align}
\zeta= \frac{  \left(M_s-\frac{1}{M_s}  \right)^2 - \frac{1(\gamma^2-1)}{\gamma}\frac{Q}{\overline{p}_1\overline{v}_1} }{(\gamma-1)^2 M_s^2}
\end{align}
The solutions are again identical with those for a perfect gas.  These solutions are shown in Fig.\ \ref{fig1} for detonations and deflagrations.  These are shown separately, for clarity.  These are discussed at length in the literature, see for example Lee's monograph on detonations \cite{lee2008detonation} or by Landau and Lifshitz \cite{landau}.  Briefly, it shows that the pressure and volume changes from the initial state 1 admit two solutions for the same wave speed, denoted as \textit{strong} or \textit{weak} for respectively the upper and lower sign in the signed expressions involving the term containing $\zeta$.  Detonations are supersonic and compressive, whether deflagrations are subsonic and expansive.  Chapman-Jouguet detonations or deflagrations are obtained when the weak and strong solutions merge and the terms involving $\zeta$ vanish.  This requires $\zeta$ to vanish, yielding the two CJ reaction waves possible, detonations and deflagrations:
\begin{align}
M_{s,CJ}^2=1+\frac{\gamma^2-1}{\gamma}\frac{Q}{\overline{p}_1\overline{v}_1}\pm\sqrt{\left( \frac{\gamma^2-1}{\gamma}\frac{Q}{\overline{p}_1\overline{v}_1}+1 \right)^2    -1}
\end{align}

These are again the same expressions as for a perfect gas, with $\overline{p}$ replacing $p$ and $\overline{v}$ replacing $v$ and the appropriate expression for the sound speed entering the definition of the Mach number.  Other jump relations are easily formulated as for the inert shocks. It can be shown that the Mach number of the flow in the frame of reference of the wave, i.e., $(D-u_2)/c_2$ is unity for the Chapman-Jouguet detonations and deflagrations.

\section{Conclusions}
We have derived the main results required for analytical work of inert and reactive gasdynamics in a medium approximated by the NASG equation of state.  The expressions obtained are transparent generalizations of the perfect gas relations and can be useful in  the modelling of compressible flows in inert and reactive dense gases, liquids and solids.    
\bibliography{references}
\bibliographystyle{elsarticle-num}
\end{document}